\documentclass[12pt,a4paper,notitlepage]{article}
\usepackage{amssymb}
\usepackage{amsmath}
\usepackage[top=1in, bottom=1.25in, left=1in, right=1in,a4paper]{geometry}
\usepackage[onehalfspacing]{setspace}
\usepackage{amsfonts}
\usepackage{amstext}
\usepackage{amsthm}
\usepackage[round]{natbib}
\usepackage{hyperref}
\usepackage{graphicx}

\numberwithin{equation}{section}

\theoremstyle{remark}
\newtheorem{remark}{Remark}
\input{tcilatex}
\begin{document}

\title{Exact computation of GMM estimators for instrumental variable
quantile regression models\thanks{%
This work was supported in part by the European Research Council
(ERC-2014-CoG-646917-ROMIA) and by the British Academy (International
Partnership and Mobility Scheme Grant, reference number PM140162).}}
\author{Le-Yu Chen\thanks{%
E-mail: lychen@econ.sinica.edu.tw} \\
{\small {Institute of Economics, Academia Sinica}} \and Sokbae Lee\thanks{%
E-mail: sl3841@columbia.edu} \\
{\small {Department of Economics, Columbia University}}\\
{\small {Centre for Microdata Methods and Practice, Institute for Fiscal
Studies} }}
\date{March 2017}
\maketitle

\begin{abstract}
We show that the generalized method of moments (GMM) estimation problem in
instrumental variable quantile regression (IVQR) models can be equivalently
formulated as a mixed integer quadratic programming problem. This enables
exact computation of the GMM estimators for the IVQR models. We illustrate
the usefulness of our algorithm via Monte Carlo experiments and an
application to demand for fish.

\medskip


\medskip

\end{abstract}

\newpage

\section{Introduction\label{Introduction}}

The instrumental variable quantile regression (IVQR) and related models have
been increasingly popular for studying the impacts of possibly endogenous
covariates on the distribution of the outcome of interest. See a recent
review by \citet{ChernozhukovHansen2013} and references therein for the
latest developments in identification, estimation, and inference as well as
the list of empirical applications.

The IVQR model admits conditional moment restrictions which can be used to
construct the estimating equations for the GMM estimation of the model
parameters. However, the sample counterparts of the IVQR estimating
equations are discontinuous in the parameters so that the resulting GMM
estimation problem becomes a non-convex and computationally non-trivial
optimization problem. \citet{honore2004} provided a heuristic for computing
the IVQR GMM estimates. \citet{ChernozhukovHansen2006} developed the inverse
quantile regression (QR) estimator that is not directly a GMM estimator but
can be shown to be asymptotically equivalent to the IVQR GMM estimator. %
\citet{Xu2017} proposed an alternative algorithm for computing the inverse
QR estimator. The Markov chain Monte Carlo (MCMC) based Laplace type
estimator of \citet{ChernozhukovHong2003} can also be used as an
approximation of the IVQR GMM estimator but it requires careful tunning in
the MCMC implementation. \citet{kaplan2015} proposed a smoothed estimating
equation approach which facilitates the GMM computation problem but requires
the choice of the smoothing parameter.

In this paper, we are concerned with exact computation of the GMM estimates
of the IVQR parameters. As pointed out by \citet{andrews1997}, heuristic
algorithms for computation of GMM estimates that do not guarantee to find
the exact global optimum or a specific level of approximation to the global
optimum may result in extremum estimators which could exhibit statistical
behavior that is quite different from that established by theory. This
source of computational uncertainty may impact on the empirical results.
Hence, as a complement to the previous work on the IVQR computation, our
paper provides a method for exact computation of the IVQR estimates within
the classical GMM framework.

Our computational algorithm is based on the method of mixed integer
optimization (MIO). Specifically, we show that the IVQR GMM estimation
problem can be equivalently formulated as a mixed integer quadratic
programming (MIQP) problem. Thanks to the developments in MIO solution
algorithms and fast computing environments, this reformulation allows us to
solve for the exact GMM estimates by using the modern efficient MIO solvers.
Well-known numerical solvers such as CPLEX and Gurobi can be used to
effectively solve large-scale MIQP problems. See 
\citet[Section
2.1]{bertsimas2016} for discussions on computational advances in solving the
MIO problems. See also \citet{Florios:Skouras:08}, %
\citet{Kitagawa:Tetenov:2015}, \citet{bertsimas2016}, and \citet{ChenLee2016}
for related but distinct work on solving non-convex optimization problems in
statistics and econometrics via the MIO approach.

The rest of this paper is organized as follows. In Section \ref{The IVQR
model}, we summarize the setup of the IVQR model and the inverse quantile
regression method of \citet{ChernozhukovHansen2006}. In Section \ref{GMM
IVQR}, we present the MIQP formulation of the IVQR GMM estimation problem.
We conduct a simulation study of the performance of the MIQP based GMM
estimates in Section \ref{Simulation} and illustrate the application of our
computation approach in a real data exercise concerning the demand
estimation in Section \ref{Empirical example}. We then conclude the paper in
Section \ref{Conclusions}.

\section{The instrumental variable quantile regression model\label{The IVQR
model}}

Let $Y$ be an outcome of interest. We consider the quantile regression model
under endogeneity, which is characterized by the structural equation 
\begin{equation}
Y=W^{\prime }\theta (U),  \label{model}
\end{equation}%
where $U$ is an unobserved scalar random variable and $W=(D,X)$ is a vector
of covariates. The covariates $D$ may not be independent of $U.$ We assume
that there is a vector of instrumental variables, denoted as $Z$, which can
be excluded from (\ref{model}) but can influence the endogenous variables $D$
such that $\dim (Z)\geq \dim (D)$ and 
\begin{equation*}
U|X,Z\sim \mathrm{Uniform}\left( 0,1\right) .
\end{equation*}%
Assume that the function $\theta (\cdot )$ in (\ref{model}) is a measurable
function such that the mapping $\tau \mapsto W^{\prime }\theta (\tau )$ is
strictly increasing in $\tau $ for almost every realization of $W$. Under
these assumptions, it follows that%
\begin{equation}
P\left( Y\leq W^{\prime }\theta (\tau )|X,Z\right) =P\left( U\leq \tau
|X,Z\right) =\tau .  \label{conditional quantile restriction}
\end{equation}%
Given a random sample, $\left( Y_{i},W_{i},Z_{i}\right) _{i=1}^{n}$ of $n$
observations, we are interested in the estimation of $\theta (\tau )$ for
some fixed values of $\tau \in (0,1)$.

The model set forth so far is the well known linear IVQR model which has
been studied by 
\citet{ChernozhukovHansen2004, ChernozhukovHansen2005,
ChernozhukovHansen2006, ChernozhukovHansen2008}, %
\citet{ChernozhukovHansenJansson2007,ChernozhukovHansenJansson2009}, and %
\citet{kaplan2015} among many others. Note that, when there is no endogenous
covariate, this model reduces to the conventional linear quantile regression
model of \citet{Koenker1978} where $W=X=Z$. In the presence of endogeneity, %
\citet{ChernozhukovHansen2005} provided further modeling assumptions such
that the quantile-specific parameter vector $\theta (\tau )$ can be causally
interpreted as the structural quantile effect in the setting with
counterfactual outcomes.

\citet{ChernozhukovHansen2006} developed primitive conditions for the
identification of $\theta (\tau )$. They also provided an inverse QR
algorithm for the estimation of $\theta (\tau )$. To describe their
algorithm, write $\theta =\left( \alpha ,\beta \right) $ such that $%
W^{\prime }\theta (\tau )=D^{\prime }\alpha \left( \tau \right) +X^{\prime
}\beta \left( \tau \right) $. Let $\Psi _{i}=\Psi \left( X_{i},Z_{i}\right) $
be a vector of transformations of instruments with $\dim (\Psi _{i})\geq
\dim \left( \alpha \right) $. Let $A$ be a given positive definite matrix.
The Chernozhukov-Hansen inverse QR procedure proceeds as follows. Let%
\begin{equation}
\widehat{\alpha }(\tau )\equiv \arg \inf\nolimits_{\alpha \in \mathcal{A}}%
\widehat{\gamma }_{\tau }\left( \alpha \right) ^{\prime }A\widehat{\gamma }%
_{\tau }\left( \alpha \right) ,  \label{outer optimization}
\end{equation}%
where 
\begin{equation}
\left( \widehat{\beta }_{\tau }\left( \alpha \right) ,\widehat{\gamma }%
_{\tau }\left( \alpha \right) \right) \equiv \arg \inf\nolimits_{\left(
\beta ,\gamma \right) \in \mathcal{B\times G}}\frac{1}{n}\sum%
\nolimits_{i=1}^{n}\rho _{\tau }\left( Y_{i}-D_{i}^{\prime }\alpha
-X_{i}^{\prime }\beta -\Psi _{i}^{\prime }\gamma \right) ,
\label{inner optimization}
\end{equation}%
$\mathcal{A}$, $\mathcal{B}$ and $\mathcal{G}$ are compact parameter spaces,
and the check function $\rho _{\tau }$ is defined by $\rho _{\tau }\left(
u\right) =u\left( \tau -1\left\{ u<0\right\} \right) $ for $u\in \mathbb{R}$%
. The inverse QR estimator is then defined by%
\begin{equation*}
\widehat{\theta }(\tau )=\left( \widehat{\alpha }(\tau ),\widehat{\beta }%
_{\tau }\left( \widehat{\alpha }(\tau )\right) \right) .
\end{equation*}%
In the procedure above, the function $\Psi $ and the matrix $A$ can vary
across $\tau $ and be replaced by their consistent estimates. Moreover, the
QR objective function can be weighted across observations. See %
\citet{ChernozhukovHansen2006} for further details.

For implementation, \citet{ChernozhukovHansen2006} proposed to solve the
outer optimization problem (\ref{outer optimization}) by the grid search
method. The inner optimization problem (\ref{inner optimization}) is the
standard quantile regression problem, which can be solved very efficiently.
Thus, when $\dim (\alpha )=1$, the inverse QR method is computationally
appealing because its implementation amounts to solving convex optimization
sub-problems within a low-dimensional global search procedure. However, this
computational merit diminishes rapidly with the increase of the number of
endogenous variables. Instead of performing grid search, \citet{Xu2017}
proposed an alternative method to compute the inverse QR estimator. Their
approach is based on exact minimization of the quadratic norm as in (\ref%
{outer optimization}) subject to the optimality conditions for the linear
programming formulation of the QR problem of (\ref{inner optimization}). %
\citet{Xu2017} showed that the resulting computational problem reduces to a
quadratic programming problem subject to complementarity constraints for
which they developed a branch-and-bound algorithm to compute the exact
solution.

\section{Exact computation of the GMM based IVQR estimator via the mixed
integer optimization approach\label{GMM IVQR}}

The conditional moment restriction (\ref{conditional quantile restriction})
can be used to form estimating equations for the GMM estimation of $\theta
\left( \tau \right) $. That is, 
\begin{equation}
E\left[ \left( 1\left\{ Y\leq W^{\prime }\theta \left( \tau \right) \right\}
-\tau \right) L\right] =0,  \label{moment restrictions}
\end{equation}%
where $L$ is a vector of instruments consisting of functions of $X$ and $Z$.
As noted by \citet{ChernozhukovHansen2006}, the inverse QR estimator, which
is not directly a GMM estimator, can be shown to be asymptotically
equivalent to the GMM estimator with the instruments $L_{\text{CH}}\equiv
\lbrack X^{\prime },\Psi \left( X,Z\right) ^{\prime }]^{\prime }.$

In this paper, we provide an algorithm for directly computing the GMM based
IVQR estimator using the orthogonality conditions (\ref{moment restrictions}%
). Let $s_{\tau }(t)$ denote the vector $(s_{\tau ,i}(t))_{i=1}^{n}$, where $%
s_{\tau ,i}(t)\equiv 1\left\{ Y_{i}\leq W_{i}^{\prime }t\right\} -\tau $ for 
$i\in \{1,...,n\}$. Let $G$ be the $n$-by-$\dim (L)$ matrix whose $i$th row
vector is $L_{i}^{\prime }$. Let $\widehat{Q}$ be a given positive definite
matrix of dimension $\dim (L)$. The GMM based IVQR estimator of $\theta
\left( \tau \right) $, denoted by $\widehat{\theta }_{GMM}\left( \tau
\right) $, is given by%
\begin{equation}
\widehat{\theta }_{GMM}\left( \tau \right) =\arg \inf \nolimits_{\theta \in {%
\Theta }} s_{\tau }\left( \theta \right) ^{\prime }G\widehat{Q}G^{\prime
}s_{\tau }\left( \theta \right) ,  \label{GMM based IVQR estimation}
\end{equation}%
where ${\Theta }$ is the compact parameter space of $\theta $.

We now present our computational algorithm, which is based on the method of
mixed integer optimization. We note that the optimization problem (\ref{GMM
based IVQR estimation}) can be equivalently formulated as the following
constrained mixed integer quadratic programming (MIQP) problem:%
\begin{eqnarray}
&&\inf_{e=\left( e_{1},...,e_{n}\right) ,\theta \in {\Theta }}\left( e-\tau
\right) ^{\prime }G\widehat{Q}G^{\prime }\left( e-\tau \right)
\label{MIO formulation of the GMM IVQR problem} \\
&&\text{s}\text{ubject to}  \notag \\
&&e_{i}\left( -M_{i}-\epsilon \right) <Y_{i}-W_{i}^{\prime }\theta \leq
\left( 1-e_{i}\right) M_{i},\text{ }i\in \{1,...,n\},
\label{sign constraints} \\
&&e_{i}\in \{0,1\},\text{ }i\in \{1,...,n\},
\label{dichotomization constraints}
\end{eqnarray}%
where $\epsilon $ is a given small and positive real scalar (e.g. $\epsilon
= 10^{-6}$ as in our simulation study), and%
\begin{equation}
M_{i}\equiv \max\nolimits_{\theta \in \Theta }\left\vert Y_{i}-W_{i}^{\prime
}\theta \right\vert ,\text{ }i\in \{1,...,n\}.  \label{Mi}
\end{equation}

We now explain the equivalence between (\ref{GMM based IVQR estimation}) and
(\ref{MIO formulation of the GMM IVQR problem}). Note that, for a given
value of $\theta \in \Theta $, the sign constraints (\ref{sign constraints})
and the dichotomization constraints (\ref{dichotomization constraints})
enforce that $e_{i}=1\{Y_{i}\leq W_{i}^{\prime }\theta \}$ for $i\in
\{1,...n\}$. Therefore, solving the constrained MIQP problem (\ref{MIO
formulation of the GMM IVQR problem}) is equivalent to solving the GMM
estimation problem (\ref{GMM based IVQR estimation}). This equivalence
enables us to employ the modern MIQP solvers to exactly compute the GMM
estimator $\widehat{\theta }_{GMM}\left( \tau \right) $. For the
implementation, note that the values $\left( M_{i}\right) _{i=1}^{n}$ in the
inequality constraints (\ref{sign constraints}) can be computed by
formulating the maximization problem in (\ref{Mi}) as linear programming
problems, which can be efficiently solved by modern optimization solvers.
Hence these values can be easily computed and stored as the input to the
MIQP formulation (\ref{MIO formulation of the GMM IVQR problem}).

\begin{remark}
Our MIQP based computational approach can be used to find the exact global
solution in the IVQR GMM estimation problem. Modern MIQP solvers employ
branch-and-bound type algorithms which maintain along the solution process
both the feasible solutions and lower bounds on the optimal objective
function value. Therefore, for computationally demanding applications, this
feature enables us to solve for an approximate IVQR GMM estimator with a
guaranteed approximation error bound, thus facilitating the design of an
early stopping rule as described in \citet[][Section 4.3]{ChenLee2016}. $\qed
$
\end{remark}

We can perform inference on $\theta \left( \tau \right) $ using the GMM
estimator $\widehat{\theta }_{GMM}\left( \tau \right) $. As noted by %
\citet{ChernozhukovHansenJansson2009}, we can take 
\begin{equation}
\widehat{Q}=\left[ \tau \left( 1-\tau \right)
n^{-1}\sum\nolimits_{i=1}^{n}L_{i}L_{i}^{\prime }\right] ^{-1}
\label{GMM weight matrix}
\end{equation}%
as a convenient and natural choice of the GMM weight matrix. By (\ref%
{conditional quantile restriction}), this weight matrix equals the inverse
of the variance of $n^{-1/2}\sum\nolimits_{i=1}^{n}s_{\tau ,i}(\theta \left(
\tau \right) )L_{i}$ conditional on $\left( L_{i}\right) _{i=1}^{n}$. Let $%
\varepsilon _{\tau }\equiv Y-W^{\prime }\theta \left( \tau \right) $. In the
GMM estimation (\ref{GMM based IVQR estimation}) with $\widehat{Q}$ given by
(\ref{GMM weight matrix}), it is straightforward to establish via empirical
process theory \citep[see e.g.,][]{pakes1989} that 
\begin{equation}
\sqrt{n}(\widehat{\theta }_{GMM}\left( \tau \right) -\theta \left( \tau
\right) )\overset{d}{\longrightarrow }N(0,\Omega ),
\label{asymptotic distribution of the GMM estimator}
\end{equation}%
where the asymptotic variance matrix $\Omega $ is given by 
\begin{equation}
\Omega =\tau \left( 1-\tau \right) \left[ \Sigma _{WL}\Sigma
_{LL}^{-1}\Sigma _{WL}^{\prime }\right] ^{-1},\Sigma _{WL}=E\left[
f_{\varepsilon _{\tau }}\left( 0|W,Z\right) WL^{\prime }\right] ,\Sigma
_{LL}=E\left[ LL^{\prime }\right] .  \label{asym. variance}
\end{equation}

We can estimate $\Sigma _{LL}$ by the sample analog $\widehat{\Sigma }%
_{LL}\equiv n^{-1}\sum\nolimits_{i=1}^{n}L_{i}L_{i}^{\prime }$. Let $%
\widehat{\varepsilon }_{\tau ,i}\equiv Y_{i}-W_{i}^{\prime }\widehat{\theta }%
_{GMM}\left( \tau \right) $. Following \citet{powell1986}, $\Sigma _{WL}$
can be consistently estimated by%
\begin{equation}
\widehat{\Sigma }_{WL}\equiv n^{-1}\sum\nolimits_{i=1}^{n}\left[ K\left( 
\widehat{\varepsilon }_{\tau ,i}/h_{n}\right) /h_{n}\right]
W_{i}L_{i}^{\prime },
\label{Powell estimator of the asymptotic variance component}
\end{equation}%
where $K\left( \cdot \right) $ is a kernel function and $h_{n}$ is a
bandwidth sequence satisfying that $h_{n}\longrightarrow 0$ and $\sqrt{n}%
h_{n}\longrightarrow \infty .$ Specific rule-of-thumb choices of $h_{n}$ can
be based on \citet{Koenker1994}. See also 
\citet[Section
3.4]{ChernozhukovHansen2006} and \citet[Section 4.4]{ChernozhukovHansen2008}
for the estimation of the IVQR variance components. Based on these results,
it is therefore straightforward to construct the confidence interval
estimates of $\theta \left( \tau \right) $ within the GMM framework.

\section{Simulation study\label{Simulation}}

In this section, we study the performance of the GMM estimator $\widehat{%
\theta }_{GMM}\left( \tau \right) $ in finite-sample simulations. We used
the MATLAB implementation of the Gurobi Optimizer (version 7.0) to solve the
MIQP problems for all numerical results of this paper.\footnote{%
The MATLAB codes for the computation of $\widehat{\theta }_{GMM}\left( \tau
\right) $ are available from the authors. This implementation requires the
Gurobi Optimizer, which is freely available for academic purposes.} All
computations were done on a desktop PC (Windows 7) equipped with 32 GB RAM
and a CPU processor (Intel i7-5930K) of 3.5 GHz.

We generated $n=100$ observations from the following simple location scale
model:%
\begin{eqnarray}
Y &=&1+D_{1}+D_{2}+D_{3}+(0.5+D_{1}+0.25D_{2}+0.15D_{3})\varepsilon ,
\label{DGP} \\
D_{1} &=&\Phi (Z_{1}+v_{1}),D_{2}=2\Phi (Z_{2}+v_{2}),D_{3}=1.5\Phi
(Z_{3}+v_{3}),  \notag
\end{eqnarray}%
where $\Phi $ denotes the cdf of the standard normal random variable, $Z_{1}$%
, $Z_{2}$ and $Z_{3}$ are independent standard normal random variables, and $%
\left( \varepsilon ,v_{1},v_{2},v_{3}\right) $ is generated independently of 
$\left( Z_{1},Z_{2},Z_{3}\right) $ from multivariate normal distribution
with mean zero and variance $0.25V$ where%
\begin{equation*}
V=\left[ 
\begin{array}{cccc}
1 & 0.4 & 0.6 & -0.2 \\ 
0.4 & 1 & 0 & 0 \\ 
0.6 & 0 & 1 & 0 \\ 
-0.2 & 0 & 0 & 1%
\end{array}%
\right] .
\end{equation*}%
By Skorohod representation, we can rewrite the model (\ref{DGP}) as%
\begin{equation*}
Y=\theta _{0}(U)+\theta _{1}(U)D_{1}+\theta _{2}(U)D_{2}+\theta _{3}(U)D_{3},
\end{equation*}%
where $U=F_{\varepsilon }\left( \varepsilon \right) $ with $F_{\varepsilon }$
being the cdf of the unobservable $\varepsilon $, and 
\begin{equation*}
\theta _{0}\left( \tau \right) =1+0.5F_{\varepsilon }^{-1}\left( \tau
\right) ,\theta _{1}\left( \tau \right) =1+F_{\varepsilon }^{-1}\left( \tau
\right) ,\theta _{2}\left( \tau \right) =1+0.25F_{\varepsilon }^{-1}\left(
\tau \right) ,\theta _{3}\left( \tau \right) =1+0.15F_{\varepsilon
}^{-1}\left( \tau \right) .
\end{equation*}

We used 500 simulation repetitions for all simulation experiments. In the
GMM estimation, we took $W=\left( 1,D_{1},D_{2},D_{3}\right) $ and $L=\left(
1,Z_{1},Z_{2},Z_{3}\right) $. The GMM weight matrix $\widehat{Q}$ was
constructed based on (\ref{GMM weight matrix}). We set the parameter space $%
\Theta $ in the MIQP problem (\ref{MIO formulation of the GMM IVQR problem})
to be the product of the intervals $[\widehat{\theta }_{j,2SLS}-10\widehat{%
\sigma }_{j,2SLS},\widehat{\theta }_{j,2SLS}+10\widehat{\sigma }_{j,2SLS}]$,
where for $j\in \{0,1,2,3\}$, $\widehat{\theta }_{j,2SLS}$ and $\widehat{%
\sigma }_{j,2SLS}$, respectively denote the parameter estimate and its
estimated heteroskedasticity-robust standard error from the two-stage least
square regression of $Y$ on the covariates $W$ using $L$ as the instruments.
The value of $\epsilon $ in (\ref{sign constraints}) was set to be $10^{-6}$.

\begin{table}[htbp]
\caption{MIQP computation time (CPU seconds)}
\label{tab01}
\begin{center}
\begin{tabular}{ccccc}
\hline\hline
$\tau $ & mean & min & median & max \\ \hline
0.25 & 94 & 37 & 92 & 197 \\ 
0.5 & 348 & 104 & 333 & 989 \\ 
0.75 & 86 & 33 & 84 & 186 \\ \hline
\end{tabular}%
\end{center}
\end{table}

We now present the simulation results. First, we report the computational
performance of our MIQP algorithm for computing the IVQR GMM estimator.
Table \ref{tab01} gives the summary statistics of the MIQP computation time
in CPU seconds across simulation repetitions. From this table, we can see
that the MIQP problems (\ref{MIO formulation of the GMM IVQR problem}) were
solved very efficiently in these simulations which incorporated three
endogenous covariates. For the two cases with $\tau \in \{0.25,0.75\}$, the
computation time was comparable. Both cases could be easily solved with the
mean and median computation time not exceeding 100 seconds and the maximum
time below 200 seconds. The case of $\tau =0.5$ appeared to be the most
computationally demanding but its maximum time remained capped within 17
minutes.

\begin{table}[h!]
\caption{Finite-sample performance of the GMM estimator}
\label{tab02}
\begin{center}
\begin{tabular}{lrrrr}
\hline\hline
& mean &  & median &  \\ 
& bias & RMSE & bias & MAE \\ \hline
$\theta _{0}\left( 0.25\right) $ & 0.0109 & 0.2436 & 0.0012 & 0.1643 \\ 
$\theta _{1}\left( 0.25\right) $ & -0.0327 & 0.3554 & -0.0048 & 0.2309 \\ 
$\theta _{2}\left( 0.25\right) $ & 0.0003 & 0.1642 & 0.0031 & 0.1008 \\ 
$\theta _{3}\left( 0.25\right) $ & 0.0064 & 0.2232 & -0.0068 & 0.1522 \\ 
\hline
$\theta _{0}\left( 0.5\right) $ & 0.0161 & 0.2498 & -0.0037 & 0.1724 \\ 
$\theta _{1}\left( 0.5\right) $ & -0.0412 & 0.3241 & -0.0316 & 0.2315 \\ 
$\theta _{2}\left( 0.5\right) $ & -0.0012 & 0.1561 & 0.0066 & 0.1038 \\ 
$\theta _{3}\left( 0.5\right) $ & 0.0012 & 0.2047 & 0.0031 & 0.1396 \\ \hline
$\theta _{0}\left( 0.75\right) $ & 0.0187 & 0.3046 & 0.0055 & 0.1849 \\ 
$\theta _{1}\left( 0.75\right) $ & -0.0358 & 0.3425 & -0.0264 & 0.2235 \\ 
$\theta _{2}\left( 0.75\right) $ & -0.0022 & 0.1820 & 0.0062 & 0.1181 \\ 
$\theta _{3}\left( 0.75\right) $ & 0.0035 & 0.2393 & -0.0016 & 0.1538 \\ 
\hline
\end{tabular}%
\end{center}
\end{table}

We now study the statistical performance of the IVQR GMM estimator. In Table %
\ref{tab02}, we report the mean and median biases, root mean squared error
(RMSE) and median absolute error (MAE) of the GMM estimators $\widehat{%
\theta }_{GMM}\left( \tau \right) $ for $\tau \in \{0.25,0.5,0.75\}$. From
these results, we find that the GMM estimators performed quite well in terms
of estimation bias. Across the three quantile cases, the estimators for $%
\theta _{1}\left( \tau \right) $ appeared to have larger dispersion in terms
of both RMSE and MAE.

\begin{table}[h!]
\caption{Comparison with asymptotic approximation}
\label{tab03}
\begin{center}
\begin{tabular}{lcc}
\hline\hline
& standard deviation & asymptotic \\ 
& in simulations & standard error \\ \hline
$\theta _{0}\left( 0.25\right) $ & 0.2434 & 0.2297 \\ 
$\theta _{1}\left( 0.25\right) $ & 0.3539 & 0.3256 \\ 
$\theta _{2}\left( 0.25\right) $ & 0.1642 & 0.1572 \\ 
$\theta _{3}\left( 0.25\right) $ & 0.2231 & 0.2059 \\ \hline
$\theta _{0}\left( 0.5\right) $ & 0.2493 & 0.2296 \\ 
$\theta _{1}\left( 0.5\right) $ & 0.3215 & 0.3049 \\ 
$\theta _{2}\left( 0.5\right) $ & 0.1561 & 0.1474 \\ 
$\theta _{3}\left( 0.5\right) $ & 0.2047 & 0.1994 \\ \hline
$\theta _{0}\left( 0.75\right) $ & 0.3040 & 0.2744 \\ 
$\theta _{1}\left( 0.75\right) $ & 0.3406 & 0.3400 \\ 
$\theta _{2}\left( 0.75\right) $ & 0.1820 & 0.1664 \\ 
$\theta _{3}\left( 0.75\right) $ & 0.2393 & 0.2283 \\ \hline
\end{tabular}%
\end{center}
\end{table}

It is also interesting to assess how well the finite-sample behavior of the
IVQR GMM estimator can be approximated by asymptotic theory. For this
purpose, our exact GMM estimator can be used to eliminate the unquantified
uncertainty on the solution inaccuracy that might emerge in a heuristic
optimization procedure. In Table \ref{tab03}, we calculated the asymptotic
standard error based on the formula (\ref{asym. variance}) evaluated at true
parameter values of the simulation design. This quantity was then compared
to standard deviation of $\widehat{\theta }_{GMM}\left( \tau \right) $ in
simulations. The results of Table \ref{tab03} indicate that the
finite-sample standard error of the GMM estimator in this simulation setup,
though being slightly larger, can be well approximated by the asymptotic
standard error.

\begin{table}[h!]
\caption{Coverage probabilities (95\% CI)}
\label{tab04}
\begin{center}
\begin{tabular}{lccc}
\hline\hline
& $0.8h_{n,HS}$ & $h_{n,HS}$ & $1.2h_{n,HS}$ \\ \hline
$\theta _{0}\left( 0.25\right) $ & 0.930 & 0.940 & 0.952 \\ 
$\theta _{1}\left( 0.25\right) $ & 0.906 & 0.914 & 0.918 \\ 
$\theta _{2}\left( 0.25\right) $ & 0.912 & 0.924 & 0.934 \\ 
$\theta _{3}\left( 0.25\right) $ & 0.916 & 0.926 & 0.938 \\ \hline
$\theta _{0}\left( 0.5\right) $ & 0.936 & 0.944 & 0.950 \\ 
$\theta _{1}\left( 0.5\right) $ & 0.938 & 0.944 & 0.952 \\ 
$\theta _{2}\left( 0.5\right) $ & 0.950 & 0.958 & 0.966 \\ 
$\theta _{3}\left( 0.5\right) $ & 0.896 & 0.916 & 0.928 \\ \hline
$\theta _{0}\left( 0.75\right) $ & 0.896 & 0.916 & 0.928 \\ 
$\theta _{1}\left( 0.75\right) $ & 0.922 & 0.938 & 0.944 \\ 
$\theta _{2}\left( 0.75\right) $ & 0.892 & 0.896 & 0.908 \\ 
$\theta _{3}\left( 0.75\right) $ & 0.918 & 0.928 & 0.942 \\ \hline
\end{tabular}%
\end{center}
\end{table}

In practice, for carrying out inference, the asymptotic variance of the GMM
estimator has to be estimated. We used the Gaussian kernel in the estimation
of $\Sigma _{WL}$. The bandwidth sequence $h_{n}$ in (\ref{Powell estimator
of the asymptotic variance component}) was based on the Hall-Sheather
bandwidth choice, which was suggested by \citet{Koenker1994} and also used
by \citet{ChernozhukovHansenJansson2009}. We also checked the sensitivity of
the inference results with respect to this bandwidth choice. Specifically,
we reported in Table 4 the finite-sample cover probabilities of the 95\%
confidence interval (CI) estimates for $\theta \left( \tau \right) $, which
were constructed based on the normal approximation theory described in
Section \ref{GMM IVQR} with three different bandwidth choices: $h_{n}\in
\{0.8h_{n,HS},h_{n,HS},1.2h_{n,HS}\}$, where $h_{n,HS}$ denotes the
Hall-Sheather bandwidth sequence. From Table 4, we find that the coverage
probabilities results were not very sensitive across bandwidth values
although the CI estimates were slightly under-sized. We also notice that the
CI estimates based on taking $h_{n}=h_{n,HS}$ or $h_{n}=1.2h_{n,HS}$
performed quite well in terms of overall performance.

\section{An illustrative empirical example: estimating the demand for fish 
\label{Empirical example}}

We illustrate usefulness of our method for exact computation of the IVQR GMM
estimator in an empirical study of the demand for fish. We used the dataset
constructed by \citet{Graddy1995} on the transactions of whiting in the
Fulton fish market in New York. The data were also previously studied in %
\citet{ChernozhukovHansen2008} and \citet{ChernozhukovHansenJansson2009} to
illustrate the econometric methods developed for quantile regression models
with endogeneity. In what follows, we mainly focused on analyzing the
results estimated by the MIQP approach and comparing them to the inverse QR
estimation results.

The data consist of 111 observations on the price and quantity of whiting
transactions aggregated by day. The outcome variable $Y$ is the logarithm of
total amount of whitings sold on each day and the endogenous explanatory
variable $D$ is the logarithm of the average daily price. The exogenous
explanatory variables include the indicators ($Monday$, $Tuesday$, $%
Wednesday $ and $Thursday$) for days of the week. The instrumental variables
are indicators ($Stormy$ and $Mixed$) for weather conditions at sea. These
instruments capture the wave height and wind speed, which should affect the
supplied quantity of fish and hence the price in the market but should not
influence the demand for fish. See \citet{Graddy1995, Graddy2006} for
further details on the operation of the Fulton fish market, and the data and
variables used for this study.

Following \citet{ChernozhukovHansenJansson2009}, we considered the simple
demand equation%
\begin{equation}
Y=\theta _{0}\left( U\right) +\theta _{1}\left( U\right) D
\label{demand equation specification}
\end{equation}%
for the estimation of $\theta _{1}$, the price elasticity of the demand,
which may vary across the demand level $U$. We also augmented the
specification (\ref{demand equation specification}) by incorporating the day
effect variables as additional controls, and then performed the estimation.
Table 5 presents the estimation results for $\theta _{1}\left( \tau \right) $
under these two different specifications. For GMM estimation results, we
took $L=\left( 1,Stormy,Mixed\right) $ as instruments and configurated the
MIQP setting in the same fashion as in Section \ref{Simulation}. We used the
Gaussian kernel and the Hall-Sheather bandwidth choice for estimating the
standard deviation of the GMM estimator and constructing the 95\% CI for $%
\theta _{1}\left( \tau \right) $. We also performed some sensitivity check
and found that the results were not very sensitive to the bandwidth choice.
Moreover, we also extracted the inverse QR and the corresponding 95\%
asymptotic CI estimation results provided by 
\citet[][Table
1]{ChernozhukovHansenJansson2009} on the same estimating model
specifications and listed them in Table \ref{tab05} for comparison.

\begin{table}[h!]
\caption{IVQR estimation of demand elasticity}
\label{tab05}
\begin{center}
\begin{tabular}{lccc}
\hline\hline
& $\tau =0.25$ & $\tau =0.5$ & $\tau =0.75$ \\ \hline
\multicolumn{4}{l}{\textit{Specification (\ref{demand equation specification}%
)}} \\ 
&  &  &  \\ 
\multicolumn{4}{l}{Estimation method: GMM via the MIQP implementation} \\ 
$\widehat{\theta }_{1}(\tau )$ & -1.0880 & -0.8876 & -0.9755 \\ 
std. dev. & 0.4773 & 0.5056 & 0.3027 \\ 
95\% CI & $\left( -2.0234,-0.1525\right) $ & $\left( -1.8787,0.1034\right) $
& $\left( -1.5689,-0.3822\right) $ \\ 
&  &  &  \\ 
\multicolumn{4}{l}{Estimation method: Inverse QR} \\ 
$\widehat{\theta }_{1}(\tau )$ & -1.3680 & -0.8860 & -1.2685 \\ 
std. dev. & 0.5704 & 0.4673 & 0.3911 \\ 
95\% CI & $\left( -2.486,-0.250\right) $ & $\left( -1.802,0.030\right) $ & $%
\left( -2.035,-0.502\right) $ \\ \hline
&  &  &  \\ 
\multicolumn{4}{l}{\textit{Specification (\ref{demand equation specification}%
) augmented with day fixed effects}} \\ 
&  &  &  \\ 
\multicolumn{4}{l}{Estimation method: GMM via the MIQP implementation} \\ 
$\widehat{\theta }_{1}(\tau )$ & -0.6915 & -0.7152 & -1.0904 \\ 
std. dev. & 0.3253 & 0.4828 & 0.2465 \\ 
95\% CI & $\left( -1.3290,-0.0540\right) $ & $\left( -1.6616,0.2312\right) $
& $\left( -1.5735,-0.6074\right) $ \\ 
&  &  &  \\ 
\multicolumn{4}{l}{Estimation method: Inverse QR} \\ 
$\widehat{\theta }_{1}(\tau )$ & -1.3635 & -0.5950 & -1.1790 \\ 
std. dev. & 0.5304 & 0.4398 & 0.3653 \\ 
95\% CI & $\left( -2.403,-0.324\right) $ & $\left( -1.457,0.267\right) $ & $%
\left( -1.895,-0.463\right) $ \\ \hline
\end{tabular}%
\end{center}
\end{table}

We now summarize the results in Table \ref{tab05}. First, we find that, for
both model specifications, the point estimates of the demand elasticity were
all negative but the magnitudes varied across quantile indices. Moreover,
both the GMM and inverse QR estimates of $\theta _{1}(\tau )$ were of
similar values under the basic specification (\ref{demand equation
specification}). When the day effect variables were included as additional
controls, the values of $\widehat{\theta }_{1}(\tau )$ across these two
estimation methods differed to a larger extent in the case of $\tau =0.25$.
Furthermore, we note that the CI results based on both the GMM and inverse
QR methods indicate that the negativity of $\theta _{1}\left( \tau \right) $
was significant for $\tau \in \{0.25,0.75\}$ but we could not reject the
case of $\theta _{1}\left( \tau \right) $ being zero at $\tau =0.5$.

\section{Conclusions\label{Conclusions}}

In this paper, we have proposed a mixed integer quadratic programming
approach for estimating the IVQR model within the GMM framework. One
possible application of our approach is panel data quantile regression for
group-level treatments \citep{chetverikov2016}. To deal with group-level
unobservables, the estimation procedure in \citet{chetverikov2016} consists
of group-by-group quantile regression followed by two-stage least squares.
They mention (in their footnote 10) that the latter step could be replaced
by an IV median regression, if one is willing to replace the usual
assumption that the group-level errors are uncorrelated with instruments
with median uncorrelation \citep{komarova2012}. This alternative step can be
computed using our computation algorithm. It is an interesting topic for
future research to fully develop this alternative to IV quantile regression
for group-level treatments.

Our approach is limited to  GMM estimators for parametric IVQR models. One may consider semiparametric models with endogeneity.  For example, \citet{ChenLintonVanKeilegom2003} considered partially linear median regression 
with some endogenous regressors as one of their examples. 
Their proposed estimator consists of a two-step procedure: in the first step, nonparametric median regression is carried out given the parameter of interest and in the second step, GMM estimation is implemented with the first step estimates as inputs. Our proposed algorithm is not directly applicable because of the first nonparametric step. It is another interesting topic
for future research to develop an algorithm to compute this kind of  two-step semiparametric quantile IV estimators.

\bibliographystyle{econometrica}
\bibliography{IVQR}

\end{document}